# A Cache-based Optimizer for Querying Enhanced Knowledge Bases

Wei Emma Zhang, *Macquarie University, Australia*
Quan Z. Sheng, *Macquarie University, Australia*
Schahram Dustdar, *TU Wien, Austria*

**Abstract**

With recent emerging technologies such as the Internet of Things (IoT), information collection on our physical world and environment can be achieved at a much higher granularity and such detailed knowledge will play a critical role in improving the productivity, operational effectiveness, decision making, and in identifying new business models for economic growth. Efficient discovery and querying such knowledge remains a key challenge due to the limited capability and high latency of connections to the interfaces of knowledge bases, e.g., the SPARQL endpoints. In this article, we present a querying system on SPARQL endpoints for knowledge bases that performs queries faster than the state-of-the-art systems. Our system features a cache-based optimization scheme to improve querying performance by prefetching and caching the results of predicted potential queries. The evaluations on query sets from SPARQL endpoints of DBpedia and Linked GeoData showcase the effectiveness of our approach.

## 1. Introduction

With recent advances in radio-frequency identification (RFID), low-cost wireless sensor devices, and Web technologies, it is possible to connect everyday objects to the Internet and to enable communications/interactions between these objects [1, 9, 10, 13]. An object (or a thing) can be an intelligent watch worn by a person that alerts her when the heart rate is too high while she is doing exercise, and a bridge equipped with sensors to monitor the stresses. Things can talk wirelessly among themselves and perform tasks that realize the machine-to-machine, human-to-machine, and machine-to-human communications [9].

Cisco predicts that there will be over 50 billion connected devices by 2020 (http://readwrite.com/2011/07/17/cisco_50_billion_things_on_the_internet_by_2020/). These devices are diverse, distributed, heterogeneous, and will generate/update a large amount of data in real time, resulting in challenging issues such as interoperability and scalability. Semantic Web technologies, which aim for a machine readable and interpretable world, can be the solutions of solving these issues [1, 9, 10]. For example, using the Resource Description Framework (RDF) data representation model and RDF Query Language (SPARQL) helps the representation and querying of the devices; shared ontologies can be used to model these devices and further improves the understanding of them; OWL axioms are used to reason and infer implicit knowledge in the data.

Instead of requiring low-level sensory data from these devices (e.g., noise level <25dBA), humans and high-level applications are often interested in high-level abstractions and perceptions that provide meanings and insights (e.g., the park is quiet) from the underlying raw sensory data. Such high-level abstractions and perceptions can then be transformed into actionable knowledge (in the form of *knowledge bases*) that helps people and applications make intelligent decisions and better respond to their environments. Knowledge bases typically exploit RDF as the data representation model. Fundamentally, RDF makes statement resources in the form of triples <subject, predicate, object>. RDF data is machine readable and allows the sharing and reuse of data across boundaries. SPARQL is a SQL-like structured query language of RDF knowledge bases. Although many efforts exist in the Semantic Web community to facilitate the SPARQL querying on knowledge bases [2, 3, 4], efficiently discovering and querying knowledge bases still remains a key challenge due to limited capability and high latency of connections to the endpoints [15].

This article makes contribution towards facilitating efficient querying on knowledge bases by leveraging concepts from the Semantic Web and machine learning techniques. Particularly, we focus on accelerating querying speed on the SPARQL endpoints. The SPARQL endpoints are interfaces that enable users to query publicly accessible knowledge bases. In this article, we present a querying system over SPARQL endpoints of knowledge bases (Section 2). Based on a prefetching and caching scheme, our system aims at accelerating the overall query response speed. We utilize a suboptimal graph edit distance function to estimate the similarity of SPARQL queries and develop an approach to transform the SPARQL queries into feature vectors. Using these feature vectors, machine learning algorithms are exploited to identify similar queries that could potentially be the subsequent queries. We then prefetch and cache the results of these queries for improving the overall querying performance. We also adopt a forecast-based cache replacement algorithm, namely Modified Simple Exponential Smoothing, to maintain only popular queries in the cache (Section 3). Our approach has been evaluated using a large set of real world SPARQL queries. The empirical results show that our approach has great potential to accelerate the querying speed on SPARQL endpoints (Section 4). We finally review the relevant literature in Section 5 and discuss some future research directions in Section 6.

## 2. Architecture Design for Querying Semantic Knowledge Bases

Before introducing the cache-based optimization approach, we briefly discuss our proposed architecture for querying the knowledge bases over the Internet of Things. The rationale is that it helps the understanding of techniques and solutions applied in querying over knowledge bases and showcase the importance of querying optimization. Figure 1 shows the layered architecture of the system. The architecture consists of two major processes of querying against SPARQL endpoints, namely the *querying process* and the *knowledge extraction process*. The querying process includes the Querying Layer and the Knowledge Bases Layer, while the knowledge extraction process involves the Data Management Layer, the Knowledge Extraction Layer, and the Knowledge Base Layer. In the following, we briefly describe the main modules of this architecture.

*Data Management Layer*. This layer manages sensory data collected from the physical things. In particular, the data management layer i) collects raw sensory data, ii) processes (e.g., filtering and cleaning) the collected data, and iii) transforms the data into meaningful information (including the integration of diverse data types into a uniform data format) that can be used in the Knowledge Extraction Layer.

*Knowledge Extraction Layer.* This layer has two main tasks: i) extracting knowledge and ii) modelling knowledge. Based on the sources of data, which can be either curated knowledge corpus or open Web resources [14], we provide two ways in extracting knowledge, namely *Curated Knowledge Extraction* (not in a real-time manner, but depending on the update of the knowledge corpus) and *Open Knowledge Extraction* (from either arbitrary text available on the Web or real-time sensory data). Curated knowledge extraction obtains facts and relationships from structured data of knowledge corpus. For instance, DBpedia (http://wiki.dbpedia.org/) extracts the structured information embedded in the Wikipedia (https://www.wikipedia.org/) articles. Open knowledge extraction leverages Open Information Extraction (Open IE) techniques to extract the meaningful information from sentences. Besides that, our system leverages the Linked Stream Data techniques [7] to compose the facts from IoT real-time data and models them in the form of relational triples.

*Knowledge Bases (KBs) and SPARQL Endpoints.* A KB represents facts and relationships associated with the logical assertions of the world. Linked Open Data (http://lod-cloud.net/) connects open KBs which are in Linked Data format for knowledge sharing and cross-domain querying. Many KBs provide the HTTP bindings for the clients to query the underlying KBs with Web interfaces. The SPARQL protocol is written in Web Services Description Language (WSDL, https://www.w3.org/TR/rdf-sparql-protocol/). In this way, clients do not need to access the backend databases or files that hold the knowledge. Instead, querying is like performing queries on a search engine. For federate queries, which aim to access data across SPARQL endpoints, SPARQL 1.1 specification introduces the SERVICE keyword for this purpose.

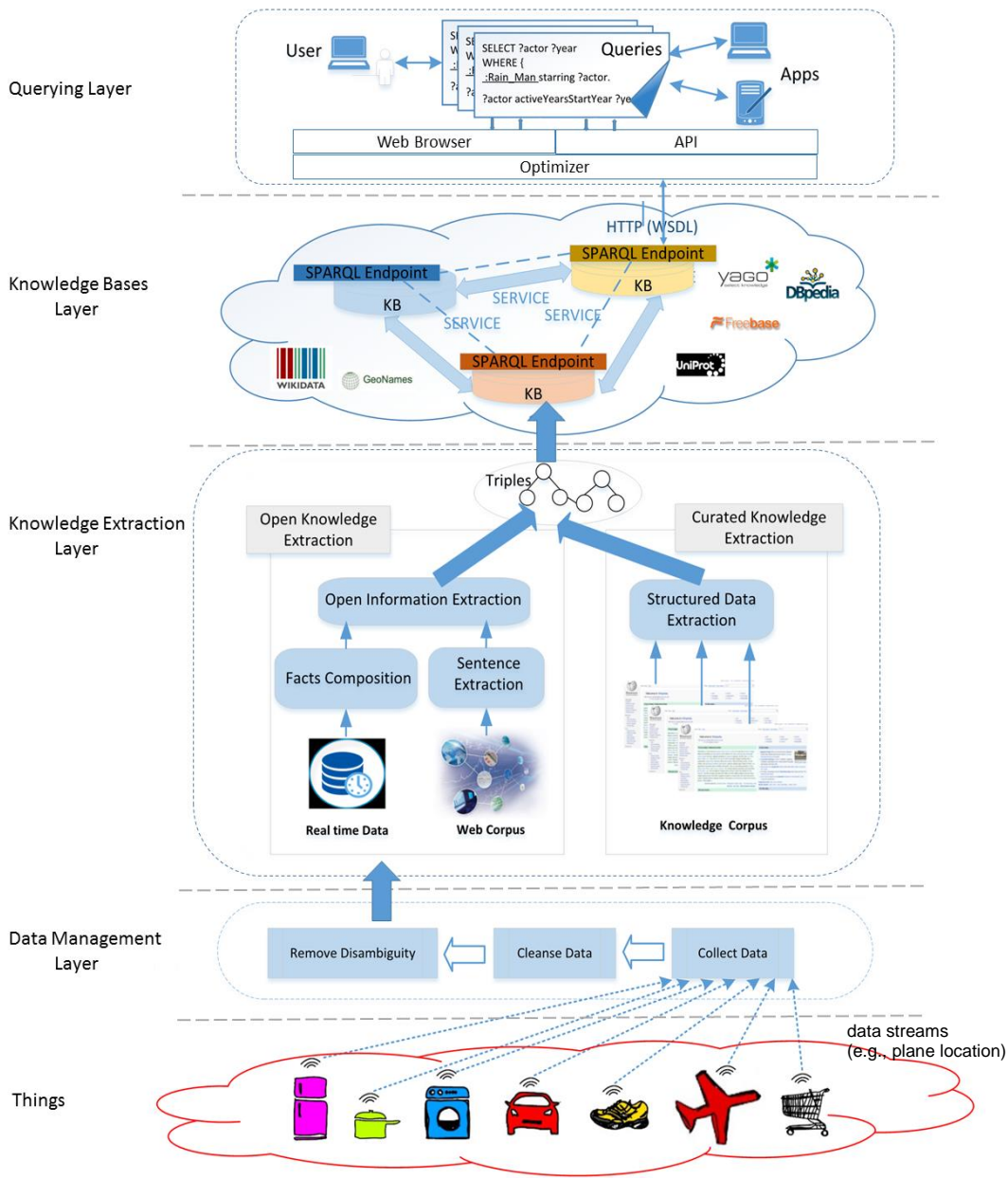

Figure 1: The System Architecture

*Optimizer.* The Optimizer in the Querying Layer aims to optimize the queries against SPARQL endpoints. We propose a caching scheme as the optimizer, which works as a proxy between clients and SPARQL endpoints. The caching scheme speeds up the overall querying process because queries are often performed more than once. Our optimizer can be either deployed as a Web browser plugin or an API, or embedded within SPARQL endpoints that act as clients to other SPARQL endpoints by interpreting the SERVICE keyword for SPARQL 1.1 federated queries. The optimizer performs three main tasks: i) it caches the queries associated with their results for future queries, ii) it controls the cache replacement to remove obsolete data, and iii) it manages the process of query answer retrieval from the cache or directly through the SPARQL endpoints. The details are discussed in Section 3.

## 3. Cache-Based Querying Optimization

Our cache-based optimizer adopts the client-side caching idea and is a domain-independent client-side caching approach. It caches the (query, result) pairs for current query and its similar queries. These similar queries are potential queries issued subsequently, so that the results can be returned directly without fetching results from the knowledge base. This will reduce the overall querying time. Our approach is motivated by the observation that end users who consume RDF-modelled knowledge typically use programmatic query clients, e.g., software or services to retrieve information from SPARQL endpoints [16]. These queries usually have repetitive query patterns and only differ in specific elements of a triple pattern (a triple pattern is a 3-tuple consisting of subject, predicate and object and each of which can be a variable). Moreover, they are usually issued subsequently. By considering these observations, we propose to prefetch and cache the query results of similar queries in advance. The problem then turns into four sub-problems and we provide our solutions correspondingly: i) we measure the similarity of SPARQL queries by considering their graph structures, ii) we predict the potential subsequent queries by training historical queries, iii) we prefetch and cache the results of potential queries, and iv) we design a smoothing-based cache replacement algorithm for record-based caching. We discuss more in detail in the following sections.

### 3.1 Suggesting Similar Queries

In order to leverage machine learning techniques to find similar queries, we firstly transfer each query to a feature vector. Then we use regression models with *Euclidean* distances between the feature vectors to suggest similar queries of $q$.

### 3.2.1 Mapping Queries to Graphs

We propose to model a SPARQL query $q$ as a graph by considering the Basic Graph Patterns (BGPs) of $q$. The BGPs of $q$ consist of triple patterns. We map all 8 different types of triple patterns to 8 structurally graphs, as shown in Figure 2(a). The eight triple patterns come from the combination of three components of triple. The black circles denote conjunction nodes. To connect the triple pattern graphs, we follow the hierarchy in the query and add a parent node to connect them. Different to graph encoding approaches using signature [8] or canonical labelling [6] which consider values of triple components, our graph mapping approach focuses on the structure of triple patterns. There are various ways to map triple patterns to graphs. However, in our work, the way that we choose for the mapping does not affect the final cache result very much. The reason is that in our work, similar queries that can lead to a cache hit are mostly the ones that are structurally the same with the current processing query.

### 3.2.2 Mapping Query Graphs to Vectors

After the graph modelling, we leverage Graph Edit Distance (GED, the minimum amount of edit operations, i.e., deletion, insertion and substitutions of nodes and edges) as the distance function to obtain query distances. However, if we use GED to find similar queries of $q$, it is extremely time consuming as the calculation of distances between $q$ and each query in training set requires large amount of computation. Therefore, we choose to construct feature vectors for $q$ using GEDs, which largely reduce the computation cost. Specifically, we use some representative queries and only calculate the distance between them and q, thus we can get a low-dimensional feature vector for $q$. The targe queries used are generated from 18 valid templates out of 25 templates in the DBPSB benchmark (we excluded queries which do not return any results: Query 1, 2, 3, 10, 16, 21 and 23) and we obtain an 18-dimension feature vector for $q$. We name this method *Template-based Feature Modelling*. The reason to use DBPSB templates is that they represent most frequently issued SPARQL query patterns. Figure 2(b) depicts this process, in which the vector [$d_1$, $d_2$... $d_{18}$] is the feature vector of $q$.

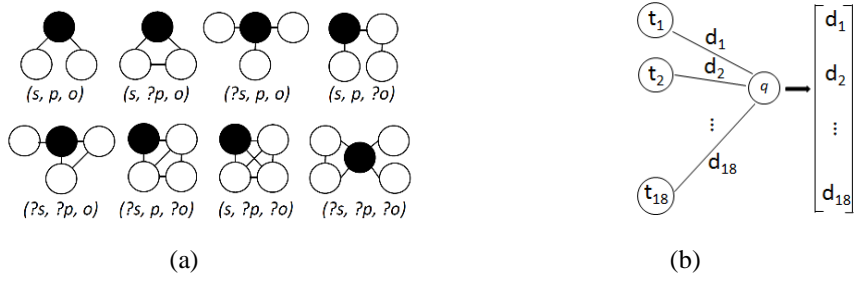

Figure 2: (a) Mapping Triple Patterns to Graphs and (b) Feature Modelling Process

### 3.2.3 Suggesting Similar Queries

After the feature vectors are obtained, we train a suggestion model with the feature vectors of training queries. We select KNN as the main technique for the suggestion model because it outperforms Support Vector Machine Regression (SVR) [2]. When training the KNN model, a K-Dimensional Tree (KD Tree) is built to compute the nearest neighbours and the distance measurement is *Euclidean* Distance. When a new query $q$ comes, we choose its K nearest neighbours as the suggested queries. More specifically, we first transform $q$ into a feature vector using the techniques from Section 3.2.1 and 3.2.2. Then we feed the vector into the trained suggestion model for similar queries recommendation.

### 3.2 Caching and Querying

The scheme consists of two main processes: i) the Querying Process and ii) the Suggestion Process. The suggestion process is a background process that will not take the resources of the querying process. In this process, similar queries to $q$ are suggested based on the training queries. The results of these similar queries are prefetched and stored in the cache. It should be noted that training queries are part of historical queries issued by the user and had been trained ahead of the querying and suggestion process. In the querying process, when a new query $q$ is issued, the framework first checks if an identical query has been cached. If it is in the cache, its results will be fetched from cache directly and then be returned to the user. Otherwise, query results will be fetched from SPARQL Endpoints (i.e., interface of knowledge bases). The results will be returned to the user and the (query, result) pairs will be put into the cache.

As the cache has limited space, we only keep the recent most hit queries in the cache, whereas other queries in the cache are considered as obsolete and will be removed from the cache. This action is cache replacement. We use a frequency based algorithm to realize the cache replacement. To be specific, the cache has an in-memory hash map where the key is the query and the value is the frequency of the query. We use Modified Simple Exponential Smoothing (MSES) [15] to estimate the hit frequency of cached queries:

$$E_t = \alpha + E_{t_{prev}} * (1 - \alpha)^{t_{prev}-t}$$

where $t_{prev}$ represents the time when a query $q$ is last hit and $E_{t_{prev}}$ denotes the previous frequency estimation for $q$ at $t_{prev}$. $\alpha$ is a smoothing constant with value between 0 and 1. This method assigns decaying weights to earlier hit queries. We perform cache replacement based on the estimation score calculated by MSES. When the top $H$ estimations are changed, the cache will be updated to reflect the new top $H$ queries and the lower ranked queries will be removed from the cache. MSES is simple but very effective—it theoretically and empirically outperforms the most used cache replacement algorithm [15]. The other reason is that traditional page-based cache replacement algorithms are not suitable for our record based cache and MSES is record based cache replacement algorithm.

## 4. Evaluation

### 4.1. Experimental Setup

The dataset used in our experiments was gathered from USEWOD 2014 challenge. We analyzed the query logs from DBPedia's SPARQL endpoint (DBpedia3.9) and Linked GeoData's endpoint (LinkedGeoData). We extracted valid SELECT queries and retrieved 198,235 queries from DBpedia3.9 and 1,790,047 queries from LinkedGeoData. Because the time consumption of the comparing approach is tremendous, we randomly chose 21,600 training queries and 5,400 test queries from the two query sets separately for all the experiments for consistency. Note that we kept the order of the queries according to their timestamps. We set up our own SPARQL Endpoint by installing local Virtuoso server and loading datasets into the Virtuoso. The server has the configuration of 64-bit Ubuntu 14.4 with 16GB RAM and 2.40GHz Intel Xeon E5-2630L v2 CPU.

We compared our feature modelling method with the method introduced in [2]. We named this approach as cluster-based feature modelling. This approach firstly clusters all the training queries using BGP as distance measurement. Then it takes the distance between $q$ and centre query of each cluster as the feature to form a feature vector for a query. The cluster-based feature modelling requires distance calculation between all queries. We also compared the overall query performance of our prefetch-based approach with the Adaptive SPARQL Query Cache (ASQC) introduced in [4], as it is the first and complete work of client-side caching of SPARQL queries. The results reported are from the second run of the test queries to avoid the warm-up stage of the cache.

### 4.2. Performance Study

Figure 3 (a) depict the impact of two feature modelling algorithms in terms of time consumption. Regarding the training time on training queries, the cluster-based approach requires 33,446 seconds, which is more than 9 hours for DBPedia3.9 queries, and 23,405 seconds (i.e., more than 6 hours) for LinkedGeoData queries. Our approach (i.e., template-based) significantly reduced the time to 1,109 seconds (i.e., 18.5 minutes) and 758 seconds (i.e., 12.6 minutes), respectively. In terms of the average query execution times of test queries, our template-based approach also outperforms the cluster-based approach.

Figure 3 (b) and (c) shows the hit rates with different number of clusters (5, 10, 15, 20, 30 in our experiments) and different K in K-NN (we chose 2, 5, 10, 20, 50,100 as K) on two datasets. For DBpedia 3.9 queries, we can find that Cluster 10 (C10 for short thereafter) gives the highest hit rate. For LinkedGeoData queries, the best performance is achieved by C15. The hit rate increases when the value of K increases for both DBpedia3.9 and LinkedGeoData queries.

Figure 3 (d) lists the performance comparison of our system with ASQC as well as their impact on servers that hold the knowledge base data. We selected the average hit rate, average query execution time (query time for short) and space usage as performance metrics. We also compared to the system that no cache was used. In order to access our datasets, we modified the code of ASQC (http://wiki.aksw.org/Projects/QueryCache). The reported results are given on DBpedia3.9 dataset and the cluster-based feature modelling. ASQC produced slightly lower hit rate (72.63%) than our system (76.65%). ASQC took 264 ms in average for one query and our system took 251 ms. When no cache was implemented, the average query time increased to 625 ms. We did not include prefetching time as it was in a separate thread. Space consumption values reflect how much memory the cache uses. In our approach, the total usage (before slash, i.e., 7.15MB) for caching 1,000 queries includes cached queries and answers, as well as the estimation records for cache replacement (after slash, i.e., 0.45KB). The numbers indicate that the most space are consumed by cached (query, result) pairs.

To evaluate the impact of caching on the server that is deployed knowledge bases and endpoints, we monitored the memory and CPU usage as well as I/O on the server. We captured the usage every 20 seconds

|  | Datasets | Cluster-based Feature modelling | Template-based Feature Modelling |
|---|---|---|---|
| Training Time on Training Queries | DBPedia3.9 | 33,446 sec. | 1,109 sec. |
|  | LinkedGeoData | 23,405 sec. | 758 sec. |
| Average Query Execution Time | DBPedia3.9 | 355 sec. | 251 sec. |
|  | LinkedGeoData | 234 sec. | 158 sec. |

(a)

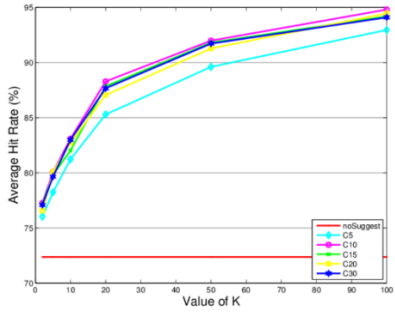

(b) Average hit rate on DBpedia3.9

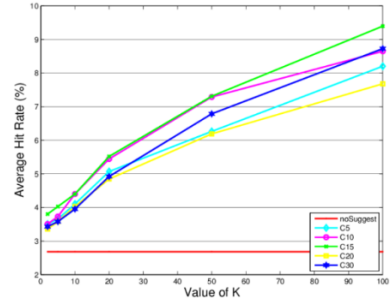

(c) Average hit rate on LinkedGeoData

|  |  | No Cache | ASQC | Our System |
|---|---|---|---|---|
| Client Side | Hit Rates | N/A | 72.63% | 76.65% |
|  | Average Query Time | 625ms | 264ms | 251ms |
|  | Space Overhead | N/A | 7.15MB | 7.15MB/0.45KB |
| Serve Side | AvgFreeMem | 224.30MB | 217.87MB | 203.74MB |
|  | AvgIO | 7.72 | 11.49 | 21.84 |
|  | AvgCPU | 9.37ms | 10.09ms | 10.68ms |

(d) Performance Comparison with ASQC

Figure 3: Time Comparison on Feature Modelling Approaches

until the querying was completed and we measured the average free memory (AvgFreeMem), average I/O (AvgFreeMem) and average CPU time (AvgCPU) including system CPU and user CPU time. We only present the results on querying DBpedia3.9 dataset due to limited space. From the result we find out that both our system and ASQC caused higher computation overhead (I/O and CPU) and memory usage on the server compared to querying without cache and ASQC performed slightly better than our system with more free memory (217.87MB vs 203.74MB), less I/O (11.49 vs 21.84) and less CPU time (9.37ms vs 10.68ms). It is because that our system requires prefetching results for similar queries from the server which leads to additional overhead. Although ASQC outperforms our system in terms of server-side overhead, our system reduces the query execution time. For large volume of queries, the saved execution time could be potentially significant.

## 5. Related Work

In this section, we review the recent representative works in two related areas, namely *Query Caching* and *Query Suggestion*.

Query caching has the rationale that it keeps the historical data for the usage of new queries. If new queries use the same data, results can be returned immediately, thereby reducing the overall query response time. Query caching was originally developed in database communities and in recent years, has been extended to triple stores that manage SPARQL queries. Martin et al. [4] first proposed caching for SPARQL queries, in which both complete triple query result and the application object are cached. However, this approach only considers repeated identical queries, while our work take both identical and similar queries into consideration. The latter ones have high potential to be requested. Yang and Wu [12] developed an approach that caches intermediate result of basic graph patterns in SPARQL queries. For a new query, the approach checks if the result of any BGP or join of BGPs of this query is cached. The hit results are joined with the other parts of the query to form the final query result, which is returned. This approach is designed to be embedded in a triple store and work with the query processing mechanism in the triple store. Papailiou et al. [6] introduced canonical labelling to identify isomorphic subgraphs in SPARQL query patterns, which are cached for subsequent querying. This solution implements a caching layer on top of the distributed partitions and dynamically updating the contents of the cache. Verborgh et al. [11] proposed a Linked Data Fragments (LDF) approach, aiming at improving data availability. It can also be regarded as a caching technique because it caches fragments of queryable data from servers that can be accessed by clients. Each client is able to process SPARQL queries on replicated fragments cached from servers.

Query suggestion is usually adopted in search engines to better understand users' information needs with the ultimate goal to improve the recall of querying. Researchers have introduced query suggestion to improve the SPARQL querying. Lehmann et al. [3] proposed to leverage a supervised learning framework to suggest SPARQL queries based on examples previously selected by users. This approach narrows the range of possible answers asked by users and requires no knowledge of the underlying schema or the SPARQL query language. Hasan [2] used a suggestion model to predict the performance of newly issued SPARQL queries. The model is trained with previously issued queries and corresponding query performance, e.g., query time. For new queries, their performances can then be predicted from the trained model. The key contribution is that the SPARQL queries are modelled as feature vectors. However, their feature modelling method is very time-consuming.

## 6. Conclusions & Research Directions

In this article, we present an optimization approach to improve the querying performance against SPARQL endpoints, the front-end of the knowledge bases. Our method transforms SPARQL queries into feature vectors. Based on these feature vectors, a learning based approach is utilized to suggest queries whose results are prefetched and cached. We leverage the graph nature of SPARQL queries and measure the distance between queries with graph edit distance. The proposed cache replacement algorithm is simple yet effective in maintaining popular queries. The evaluations on real-world queries show that our approach greatly outperforms the state-of-the-art approaches and demonstrate the potential to speed up average querying process on SPARQL endpoints.

We hope this article sheds light on the research of querying knowledge bases in a more efficient way. Future work directions include: i) more experimental studies using large-scale Web data, ii) online learning techniques to deal with the dynamicity of Web data, and iii) investigation of user behaviours that can be used to better predict future queries.